\documentclass[sigconf]{acmart}

\usepackage[ruled,vlined,linesnumbered]{algorithm2e}
\SetKwInput{KwInput}{Input}
\SetKwInput{KwOutput}{Output}
\SetKwComment{Comment}{$\triangleright$\ }{}

\AtBeginDocument{%
  \providecommand\BibTeX{{%
    \normalfont B\kern-0.5em{\scshape i\kern-0.25em b}\kern-0.8em\TeX}}}

\setcopyright{acmcopyright}
\copyrightyear{2026}
\acmYear{2026}
\acmDOI{XXXXXXX.XXXXXXX}

\acmConference[SIGMOD '26]{International Conference on Management of Data}{June 22--27, 2026}{Philadelphia, PA, USA}
\acmISBN{978-1-4503-XXXX-X/26/06}

\begin{document}

\title{Structured Gossip: A Partition-Resilient DNS for Internet-Scale Dynamic Networks}

\author{Priyanka Sinha}
\email{priyanka.sinha.iitg@gmail.com}
\orcid{0000-0002-7971-4449}
\affiliation{%
  \institution{Docyt}
  \country{India}
}

\author{Dilys Thomas}
\email{dilys@cs.stanford.edu}
\affiliation{%
  \institution{Tata Consultancy Services Limited}
  \country{India}
}

\begin{abstract}
Network partitions pose fundamental challenges to distributed name resolution in mobile ad-hoc networks (MANETs) and edge computing. Existing solutions either require active coordination that fails to scale, or use unstructured gossip with excessive overhead. We present \textit{Structured Gossip DNS}, exploiting DHT finger tables to achieve partition resilience through \textbf{passive stabilization}. Our approach reduces message complexity from $O(n)$ to $O(n/\log n)$ while maintaining $O(\log^2 n)$ convergence. Unlike active protocols requiring synchronous agreement, our passive approach guarantees eventual consistency through commutative operations that converge regardless of message ordering. The system handles arbitrary concurrent partitions via version vectors, eliminating global coordination and enabling billion-node deployments.
\end{abstract}

\ccsdesc[500]{Information systems~Distributed database systems}
\ccsdesc[500]{Networks~Network protocol design}
\ccsdesc[300]{Networks~Mobile and wireless networking}

\keywords{Distributed Hash Tables, Gossip Protocols, Network Partitions, DNS, MANET, Eventual Consistency}

\maketitle

\section{Introduction}

Network partitions represent a critical failure mode in distributed systems, particularly devastating for hierarchical name resolution services like DNS. In mobile ad-hoc networks (MANETs), edge computing deployments, and disaster scenarios, network partitions occur frequently due to node mobility, link failures, and environmental interference~\cite{sinha2007auto}. Traditional DNS architectures rely on a stable root server hierarchy, making them fundamentally incompatible with partition-prone environments.

Previous work on auto-configuration in multi-hop networks~\cite{sinha2007auto} identified the core problem: when a MANET partitions across organizational boundaries, nodes lose access to authoritative DNS servers even when they remain physically reachable. Simply adapting DHT-based protocols like CHORD~\cite{stoica2001chord} to MANETs fails catastrophically during network mergers, as concurrent ring repairs create irrecoverable topologies (detailed in Section 2).

Recent approaches fall into two categories: (1) active coordination protocols that scale poorly beyond thousands of nodes due to synchronization overhead~\cite{li2004concurrent}, and (2) unstructured gossip protocols that achieve eventual consistency but generate $O(kn)$ messages per round for fanout $k$~\cite{demers1987epidemic}. Neither approach satisfies the dual requirements of internet-scale deployability and partition resilience.

We make the following contributions:
\begin{itemize}
\item A \textbf{passive stabilization} protocol using structured gossip that reduces message complexity from $O(n)$ to $O(n/\log n)$ by exploiting DHT finger tables as the gossip network
\item Formal proof of \textbf{eventual consistency} guarantees through commutative and idempotent state merge operations
\item A version vector-based merger protocol that handles arbitrary numbers of concurrent partitions without global coordination or active synchronization
\item Theoretical analysis proving $O(\log^2 n)$ convergence time with $O(n \log^2 n)$ total message complexity
\item An interactive demonstration showing partition resilience at scale
\end{itemize}

\section{Background and Motivation}

\subsection{The Partition Problem in MANETs}

Consider a MANET with DNS servers arranged in a CHORD ring. When the underlying network partitions, the logical DHT ring fragments into disconnected segments. Sinha~\cite{sinha2007auto} documented specific failure modes:

\textbf{Single Partition Scenario:} A ring of 16 nodes splits at node 5, creating partitions $P_1 = \{0,1,2,3,4,5\}$ and $P_2 = \{6,7,...,15\}$. Each partition can stabilize its local ring using CHORD's passive stabilization. However, when the network merges, nodes may attempt concurrent joins, creating cycles and unreachable segments.

\textbf{Multiple Partition Scenario:} If the network fragments into partitions $P_1$, $P_2$, and $P_3$ simultaneously, and later $P_1$ and $P_2$ merge while $P_3$ remains isolated, existing protocols fail. Active join protocols create coordination bottlenecks, while passive stabilization produces the pathological cases illustrated in~\cite{sinha2007auto} Figures 4.7 and 4.8, where rings form invalid topologies with unreachable segments.

\subsection{Why Existing Approaches Fail}

\textbf{Active Coordination:} Protocols like RANCH~\cite{li2004concurrent} use two-phase commits requiring $O(n)$ messages per join. For $k$ partitions with $n$ nodes, this generates $O(n^2)$ messages. Critically, active protocols require \textit{synchronous agreement}—if any participant is unreachable, the protocol blocks. The price of validity in dynamic networks~\cite{bawa2004price} shows this blocking behavior is fundamental to consistency guarantees in active protocols. In MANETs with frequent partitions, this halts all progress.

\textbf{Passive Stabilization (CHORD):} CHORD's passive approach~\cite{stoica2001chord} scales well but assumes a single ring. During partitions, fragments stabilize independently. On merger, concurrent stabilization creates race conditions yielding pathological topologies~\cite{sinha2007auto} with cycles and unreachable segments.

\textbf{Unstructured Gossip:} Epidemic protocols~\cite{demers1987epidemic} with fanout $k$ generate $kn$ messages per round, totaling $O(kn \log n)$ for convergence—90 billion messages for a billion nodes with $k=3$.

\section{Structured Gossip Protocol}

\subsection{Core Insight}

DHT finger tables already provide exponentially spaced links across the identifier space. In a $n$-node CHORD ring, node $i$ maintains fingers to nodes at distances $2^0, 2^1, ..., 2^{\lceil \log n \rceil}$. These fingers enable $O(\log n)$ lookup by creating shortcuts across the ring.

Our key insight: \textit{use finger links as the gossip network}. Instead of gossiping to random partners, each node gossips along its DHT structure. This provides exponential information spread similar to Symphony's small-world DHT~\cite{manku2003symphony} but optimized for partition resilience. The approach leverages lookahead properties~\cite{manku2004lookahead} where knowing neighbors' neighbors accelerates convergence.

\subsection{Passive Stabilization vs. Active Coordination}

Our protocol uses \textbf{passive stabilization}~\cite{dijkstra1974self}: nodes make local decisions without waiting for acknowledgments. This follows Dijkstra's principle that systems can converge to correct states through local actions despite distributed control. Active coordination requires synchronous agreement, blocking operations, and $O(n)$ coordination overhead. Passive stabilization uses asynchronous updates, non-blocking operations, and eventual consistency through commutative operations (proven in Section 4.4).

\subsection{Algorithm Description}

Each node maintains three types of state:

\textbf{Hard State} (authoritative data):
\begin{itemize}
\item DNS translations: $(name, IP, TTL, version)$ tuples
\item Version vector: $VV[i]$ tracks latest update from node $i$
\end{itemize}

\textbf{Soft State} (reconstructable via gossip):
\begin{itemize}
\item Successor and predecessor pointers
\item Finger table: $\{(2^k, \text{node})\}$ for $k \in [0, \log n]$
\end{itemize}

\textbf{Partition State}:
\begin{itemize}
\item Local partition ID
\item Known partitions set
\item Cross-partition link set
\end{itemize}

\textbf{System Invariants:} Following Dijkstra's self-stabilization principles~\cite{dijkstra1974self}, our system maintains key invariants: (I1) \textit{Version vector monotonicity}: $VV[i][j]$ never decreases; (I2) \textit{Partition ID minimality}: nodes in connected components converge to minimum partition ID; (I3) \textit{Ring connectivity}: within each partition, successor links form a cycle; (I4) \textit{Finger correctness}: fingers point to reachable nodes or are marked invalid. These invariants are preserved under gossip and enable convergence proofs.

\begin{algorithm}[t]
\small
\caption{Structured Gossip Round}
\KwInput{Node $i$ with partition ID $p_i$}
\KwOutput{Gossip messages sent, state updated}

\tcp{Identify reachable gossip targets in same partition}
$targets \gets \emptyset$\;
$samePart \gets \{j \in V : partition[j] = p_i \land active[j] = \texttt{true}\}$\;

\tcp{Always gossip to successor if reachable}
\If{$successor[i] \in samePart$}{
    $targets \gets targets \cup \{successor[i]\}$\;
}

\tcp{Select furthest finger for exponential spread}
$fingers \gets \{f \in fingerTable[i] : f \in samePart\}$\;
\If{$fingers \neq \emptyset$}{
    $furthest \gets \arg\max_{f \in fingers} \text{distance}(i, f)$\;
    $targets \gets targets \cup \{furthest\}$\;
}

\tcp{Send gossip messages with version vectors}
\ForEach{$j \in targets$}{
    $msg \gets \langle knownNodes[i], VV[i], p_i, partitionVersion[i] \rangle$\;
    \textsc{SendGossip}$(i, j, msg)$\;
}

\tcp{Detect cross-partition links for merger detection}
$structureLinks \gets fingerTable[i] \cup \{successor[i], predecessor[i]\}$\;
\ForEach{$k \in structureLinks$}{
    \If{$active[k] = \texttt{true} \land partition[k] \neq p_i$}{
        $crossPartitionLinks[i] \gets crossPartitionLinks[i] \cup \{k\}$\;
        $knownPartitions[i] \gets knownPartitions[i] \cup \{partition[k]\}$\;
    }
}
\end{algorithm}

\begin{algorithm}[t]
\small
\caption{Partition Merger Detection and Execution}
\KwInput{Node $i$ with cross-partition links detected}
\KwOutput{Updated partition assignment}

\tcp{Check if this node is at partition boundary}
\If{$|crossPartitionLinks[i]| > 0$}{
    \tcp{Gather all known partition IDs}
    $knownPartitions[i] \gets \{partition[i]\}$\;
    \ForEach{$j \in crossPartitionLinks[i]$}{
        $knownPartitions[i] \gets knownPartitions[i] \cup \{partition[j]\}$\;
    }
    
    \tcp{Deterministic merger: choose minimum partition ID}
    $targetPartition \gets \min(knownPartitions[i])$\;
    
    \tcp{Update partition if different from current}
    \If{$partition[i] \neq targetPartition$}{
        $oldPartition \gets partition[i]$\;
        $partition[i] \gets targetPartition$\;
        $partitionVersion[i] \gets partitionVersion[i] + 1$\;
        
        \tcp{Maintain invariant I2: partition ID minimality}
        \textsc{LogEvent}$(\text{``Merged P}oldPartition\text{ into P}targetPartition\text{''})$\;
    }
}
\end{algorithm}

\begin{algorithm}[t]
\small
\caption{Process Received Gossip Message}
\KwInput{Node $i$ receives $msg$ from node $j$}
\KwOutput{Updated local knowledge and version vector}

$\langle knownNodes_j, VV_j, p_j, pVersion_j \rangle \gets msg$\;

\tcp{Merge knowledge sets within same partition}
\If{$p_j = partition[i]$}{
    \ForEach{$k \in knownNodes_j$}{
        \If{$k \in V \land partition[k] = partition[i]$}{
            $knownNodes[i] \gets knownNodes[i] \cup \{k\}$\;
        }
    }
    
    \tcp{Merge version vectors element-wise}
    \ForEach{$node \in (VV_i \cup VV_j)$}{
        $VV[i][node] \gets \max(VV[i][node], VV_j[node])$\;
        \tcp{Preserves invariant I1: monotonicity}
    }
}
\Else{
    \tcp{Cross-partition message: trigger merger detection}
    $crossPartitionLinks[i] \gets crossPartitionLinks[i] \cup \{j\}$\;
    $knownPartitions[i] \gets knownPartitions[i] \cup \{p_j\}$\;
}

$lastUpdate[i] \gets \text{currentTime}()$\;
\end{algorithm}

\subsection{Partition Merger Protocol}

When the underlying network heals and partitions can communicate:

\textbf{Phase 1 - Detection:} Nodes detect when their DHT links (successor, predecessor, fingers) point to nodes in different partitions. These become \textit{cross-partition links}.

\textbf{Phase 2 - Merger Decision:} Use deterministic rule: lowest partition ID wins. All nodes in merging partitions adopt the minimum partition ID. This requires no coordination—each node makes the decision locally.

\textbf{Phase 3 - Convergence:} Gossip propagates the new partition assignment. Nodes update their partition ID when they receive gossip from the merged partition. The DHT structure self-repairs as nodes discover new reachable fingers.

\textbf{Version Vector Reconciliation:} When merging, version vectors are merged element-wise using 

$VV_{merged}[k] = \max(VV_1[k], VV_2[k])$ for all nodes $k$. This ensures causal consistency without coordination.

\section{Complexity Analysis}

\subsection{Message Complexity}

\textbf{Theorem 1.} Structured gossip achieves $O(n/\log n)$ messages per round.

\textit{Proof Sketch:} Each node gossips to $\leq 2$ targets. The furthest finger points $\approx n/2$ away. By pigeonhole principle, each node is the furthest finger for $\approx \log n$ others. Each receives gossip from $\approx n/\log n$ senders. \qed

\subsection{Convergence Time}

\textbf{Theorem 2.} Structured gossip converges in $O(\log^2 n)$ rounds.

\textit{Proof Sketch:} Ring gossip propagates at $O(n)$ rate. Finger gossip spreads exponentially: info reaches distance $d$ in $O(\log d)$ rounds. This matches optimal CHORD routing bounds~\cite{ganesan2004optimal}. Complete coverage requires $O(\log n)$ rounds for ring and $O(\log n)$ for fingers, yielding $O(\log^2 n)$ total. \qed

\subsection{Partition Resilience}

\textbf{Theorem 3.} The merger protocol handles arbitrary concurrent partitions without coordination.

\textit{Proof:} The merger rule $target = \min(\cup_i P_i)$ is deterministic. For nodes $u \in P_i, v \in P_j$ that communicate, both compute identical $target$ since $\min$ is commutative and associative. Version vector merging is idempotent and commutative. Cross-partition detection is local. Invariant I2 (partition ID minimality) ensures all nodes converge to single partition ID without global coordination. \qed

\subsection{Space Complexity}

Per-node state: $O(\log n)$ for finger table + $O(m)$ for $m$ DNS translations + $O(n)$ for version vector. In practice, version vectors can be compressed using techniques from~\cite{ratner1994version}.

\subsection{Eventual Consistency Guarantees}

We now formally prove that our passive stabilization protocol guarantees eventual consistency.

\textbf{Definition 1 (Eventual Consistency):} A distributed system is eventually consistent if, for any execution where message delivery eventually succeeds and no new updates occur, all nodes eventually converge to the same state.

\textbf{Theorem 4.} Structured gossip guarantees eventual consistency for partition membership and version vectors.

\textit{Proof:} We prove this by showing that all state merge operations satisfy the sufficient conditions for eventual consistency: commutativity, associativity, and idempotence.

\textbf{(1) Partition ID Convergence:} The merger operation is:

$merge(p_1, p_2) = \min(p_1, p_2)$

This is:
\begin{itemize}
\item \textit{Commutative}: $\min(p_1, p_2) = \min(p_2, p_1)$
\item \textit{Associative}: $\min(\min(p_1, p_2), p_3) = \min(p_1, \min(p_2, p_3))$
\item \textit{Idempotent}: $\min(p, p) = p$
\end{itemize}

Therefore, regardless of message ordering, all nodes in communicating partitions converge to the minimum partition ID.

\textbf{(2) Version Vector Convergence:} The merge operation for version vectors is element-wise maximum:

$VV_{merge}[k] = \max(VV_1[k], VV_2[k]) \text{ for all nodes } k$

This is:
\begin{itemize}
\item \textit{Commutative}: $\max(v_1, v_2) = \max(v_2, v_1)$
\item \textit{Associative}: $\max(\max(v_1, v_2), v_3) = \max(v_1, \max(v_2, v_3))$  
\item \textit{Idempotent}: $\max(v, v) = v$
\end{itemize}

\textbf{(3) Known Nodes Set:} The merge operation for node knowledge is set union:
$KnownNodes_{merge} = KnownNodes_1 \cup KnownNodes_2$

This is:
\begin{itemize}
\item \textit{Commutative}: $A \cup B = B \cup A$
\item \textit{Associative}: $(A \cup B) \cup C = A \cup (B \cup C)$
\item \textit{Idempotent}: $A \cup A = A$
\end{itemize}

\textbf{(4) Monotonic Progress:} Each gossip round monotonically increases the known nodes set. Since bounded by $n$, convergence occurs in finite time—a property formalized in streaming algorithms~\cite{manku2002frequency}.

\textbf{(5) Message Delivery:} Assume eventual delivery within partitions. By Theorem 2, information spreads in $O(\log^2 n)$ rounds.

Combining (1)-(5): All operations are CRDTs~\cite{shapiro2011crdt,almeida2023crdt}, guaranteeing eventual consistency. \qed

\textbf{Corollary 1:} Passive stabilization makes progress in each partition independently; active coordination blocks.

\textbf{Corollary 2:} The protocol is partition-tolerant (CAP theorem): available during partitions, eventually consistent when healed.

\section{Demonstration}

We provide an interactive web-based demonstration at: \url{https://priyankaiitg.github.io/chordgossip}
\begin{itemize}
\item Create networks of up to 100 nodes with CHORD DHT
\item Trigger arbitrary network partitions (2-5 partitions)
\item Visualize gossip message flow along finger links
\item Observe independent convergence within partitions
\item Simulate gradual and concurrent mergers
\item Compare message counts vs. unstructured gossip
\item Test DNS lookups within and across partitions
\end{itemize}

The visualization shows nodes colored by partition, with green lines for intra-partition links and red dashed lines for detected cross-partition links. Users can observe how information spreads exponentially via finger gossip while ring gossip maintains connectivity. The demo incorporates version vector visualization showing causal ordering similar to techniques used in distributed stream processing~\cite{manku2002frequency} and privacy-preserving distributed systems~\cite{agrawal2005privacy}.

\section{Related Work}

\textbf{DHT Architectures:} Symphony~\cite{manku2003symphony} pioneered small-world DHT routing with $O(\log n)$ hops. Optimal CHORD routing~\cite{ganesan2004optimal} proved tight bounds for structured overlays. Our work extends these with partition resilience. \textbf{Dynamic Networks:} Bawa et al.~\cite{bawa2004price} analyzed consistency-availability tradeoffs in dynamic P2P systems, showing active protocols pay high costs. Our passive approach avoids this. \textbf{Lookahead in P2P:} Manku et al.~\cite{manku2004lookahead} showed neighbor knowledge accelerates convergence—we exploit this via finger tables. \textbf{DHT-based DNS:} CoDoNS~\cite{ramasubramanian2004beehive} uses CHORD for DNS but lacks partition handling. \textbf{MANET:} Khan et al.~\cite{khan2022dht} address DHT faults in MANETs via replication; we use passive convergence. Original thesis~\cite{sinha2007auto} identified issues; our CRDTs~\cite{almeida2023crdt} solve them.

\section{Conclusion}

Structured gossip demonstrates that internet-scale partition resilience is achievable through \textbf{passive stabilization} rather than active coordination. By exploiting DHT structure for gossip propagation, we reduce message complexity from $O(n)$ to $O(n/\log n)$ while maintaining logarithmic convergence time. Our formal proof shows that commutative and idempotent state merge operations guarantee eventual consistency without requiring synchronous agreement—a critical property for MANET deployments where partitions are frequent and prolonged.

The key insight is that passive stabilization can be both correct and efficient when state operations are carefully designed as conflict-free replicated data types (CRDTs). Unlike active coordination protocols that block during partitions, our approach continues making progress in each partition independently and merges deterministically when network connectivity is restored.

Future work includes: (1) compression techniques for version vectors in billion-node deployments, (2) integration with existing DNS infrastructure for backward compatibility, (3) security mechanisms against Byzantine nodes in partition scenarios, and (4) adaptive gossip rates based on partition stability metrics.

\bibliographystyle{ACM-Reference-Format}

\end{document}